# Time-multiplexed structured illumination using a DMD for optical diffraction tomography

Kyeoreh Lee,[1,2,†] Kyoohyun Kim,[1,2,†] Geon Kim,[1,2] Seungwoo Shin,[1,2] YongKeun Park[1,2,3,*]

[1]*Department of Physics, Korea Advanced Institute of Science and Technology (KAIST), Daejeon 34141, Republic of Korea*
[2]*KI for Health Science and Technology (KIHST), KAIST, Daejeon 34141, Republic of Korea*
[3]*Tomocube Inc., Daejeon 34051, Republic of Korea*
[†]*These authors contributed equally*
*Corresponding author: yk.park@kaist.ac.kr*

**We present a novel illumination control technique for optical diffraction tomography (ODT). Various spatial frequencies of beam illumination were controlled by displaying time-averaged sinusoidal patterns using a digital micromirror device (DMD). Compared to the previous method using binary Lee holograms, the present method eliminates unwanted diffracted beams which may deteriorate the image quality of the ODT. We demonstrated the capability of the present method by reconstructing three-dimensional refractive index (RI) distributions of various samples, with high RI sensitivity ($\sigma_{\Delta n}$ = 3.15 × 10⁻⁴), and reconstructing 3-D RI tomograms of biological samples, which provided quantitative biochemical and morphological information about the samples.**

Optical diffraction tomography (ODT) has emerged as an invaluable tool for measuring the three-dimensional (3-D) refractive index (RI) distribution of biological samples [1-4]. Because the RI is an intrinsic optical property of a material, ODT enables the label-free quantitative imaging of biological samples without the use of exogenous labeling agents such as fluorescent proteins or organic dyes. Recently, ODT and 3-D quantitative phase imaging techniques have been widely applied to investigate various fields of research, including hematology [5-10], immunology [11, 12], neuroscience [13, 14], infectious diseases [5, 15, 16], cancer cells [17, 18], and industrial applications [19].

As an optical analog to X-ray computed tomography, ODT measures multiple 2-D holograms of a sample from various illumination angles, and a 3-D RI tomogram is reconstructed from measured holograms via an inverse scattering algorithm. Thus, in ODT, it is crucial to achieve stable and precise control of the angle of the incident beam impinging on the sample. Various control approaches have previously been proposed, including a galvanomirror [20-22], a spatial light modulator [23], and mechanical rotation of the sample [24-26].

Recently, a digital micromirror device (DMD) was employed with ODT, and demonstrated both high-speed control and stable operation without mechanical movements [27, 28]. The approach exploits the Lee hologram method [29, 30]. To control the illumination angle, binary amplitude patterns, in which a plane wave of a specific spatial frequency is modulated, were projected on a DMD. Although it provides fast control capability, displaying a Lee hologram using the 1-bit (binary) patterns of a DMD inevitably results in unwanted diffraction noise, which significantly deteriorates the image quality of the reconstructed tomograms. Previously, to remedy such effects, an annular aperture has been used, which selects only the first-order diffracted beam and blocks all other beams. However, the use of the aperture strictly limits the range of illumination angles, so that the degrees of freedom available to control the illumination beam are limited.

In this Letter, we present a novel illumination control method for ODT that overcomes both the limitations of speed and unwanted diffraction noise. We propose and experimentally demonstrate a time-multiplexed structured illumination using a DMD. Instead of displaying a binary hologram onto a DMD, each pixel of the DMD was set to rapidly flicker and display a high-bit-depth structured illumination with a sinusoidal pattern. Using a DMD with the time-multiplexed structured illumination enables an illumination wavefront to be generated with precisely defined spatial frequencies, and removes unwanted diffraction noise. We explain the principle of the proposed method and then compare it with previous methods. In addition, using the proposed method we demonstrate 3-D RI tomogram measurements of various biological cells, including a human red blood cell and a HeLa cell.

The principle of the method can be explained in two parts: (1) grey-scaled intensity images can be projected by a DMD, a binary intensity modulator, using time multiplexing; (2) structured illumination with a sinusoidal intensity pattern is composed of three distinct spatial frequencies. In principle, the main physical concept of the present method is the same as that of structured illumination microscopy (SIM) [31]; this work can be understood as a holographic version of SIM.

A schematic diagram explaining the principle is presented in Fig. 1. Compared to the previous method which uses Lee holograms with spatial filtering [27, 28], we exploited the time-multiplexed structured illumination of sinusoidal intensity patterns. A sinusoidal intensity pattern can be composed of three plane waves [Fig.1(a)] as,

$$1+\cos(\mathbf{k}\boldsymbol{\rho}+\varphi) = e^0 + \tfrac{1}{2}e^{+i\mathbf{k}\boldsymbol{\rho}}e^{+i\varphi} + \tfrac{1}{2}e^{-i\mathbf{k}\boldsymbol{\rho}}e^{-i\varphi}, \qquad (1)$$

where $\boldsymbol{\rho}$ is a displacement vector on the DMD plane; $\mathbf{k}$ is a wavevector, which can be related to the spatial period $\Lambda$ of the sinusoidal pattern as $|\mathbf{k}| = 2\pi/\Lambda$; $\varphi$ is a phase value where $\boldsymbol{\rho} = 0$.

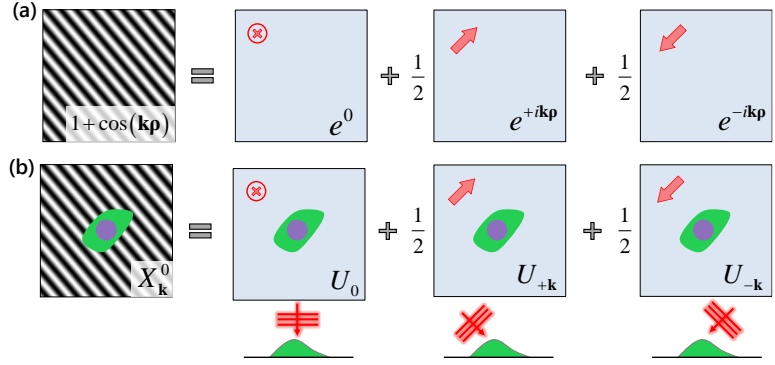

**Fig. 1**. Linear decomposition of the time-multiplexed structured illumination for optical diffraction tomography. (a) Two-dimensional sinusoidal pattern, composed of three different plane wave components, which is used for illumination. (b) the corresponding field scattered from a sample.

When a sample is illumination with the intensity pattern in Eq. 1, the transmitted light field $X_{\mathbf{k}}^{\varphi}$ can then be expressed as the superposition of scattered fields $U_k$ corresponding to plane wave illuminations $e^{i\mathbf{k}\boldsymbol{\rho}}$ [Fig. 1(b)] as,

$$X_{\mathbf{k}}^{\varphi} = U_0 + \tfrac{1}{2}U_{+\mathbf{k}}e^{+i\varphi} + \tfrac{1}{2}U_{-\mathbf{k}}e^{-i\varphi}, \qquad (2)$$

Equation (2) is a linear equation of three unknown variables ($U_0$, $U_{+\mathbf{k}}$, and $U_{-\mathbf{k}}$). Therefore, Eq. (2) can be solved with more than three measurements with various phase values $\varphi$. The control of $\varphi$ can be simply achieved by laterally shifting the sinusoidal pattern in Eq. (1), which can be understood as analogous to phase-shifting interferometry [32].

In order to generate the gray-scaled sinusoidal intensity pattern in Eq. (1), we used the built-in video mode display feature of the DMD software. The time-multiplexing display method has been widely utilized in DMD-based video projectors. In order to display an 8-bit-depth sinusoidal intensity pattern, each grey-scale intensity image is first divided into 8 binary images corresponding to $2^7$, $2^6$, ..., $2^0$ digits in binary notation. Then, each binary image is projected on a DMD with different flipping frequencies. This structured illumination impinges onto a sample, and then the transmitted light field is measured using interferometric microscopy. When measuring transmitted light fields, the exposure time of the camera was optimized to obtain the hologram information expressed in Eq. (2), which ensures the time-multiplexed structured illumination.

The setup for the time-multiplexed structured illumination using a DMD and ODT is shown in Fig. 2. A Mach-Zehnder interferometric microscope was utilized to measure optical fields [Fig. 2(a)] [27, 28]. A diode-pumped solid-state laser beam ($\lambda$ = 532 nm, 50 mW, Covolt AB, Sweden) was coupled into a 2×2 single-mode fiber optic coupler (TW560R2F2, Thorlabs, Inc., NJ, USA), which works as a spatial filter, and also splits the laser beam into two arms. One arm was used as a reference beam, and the other arm was the sample beam.

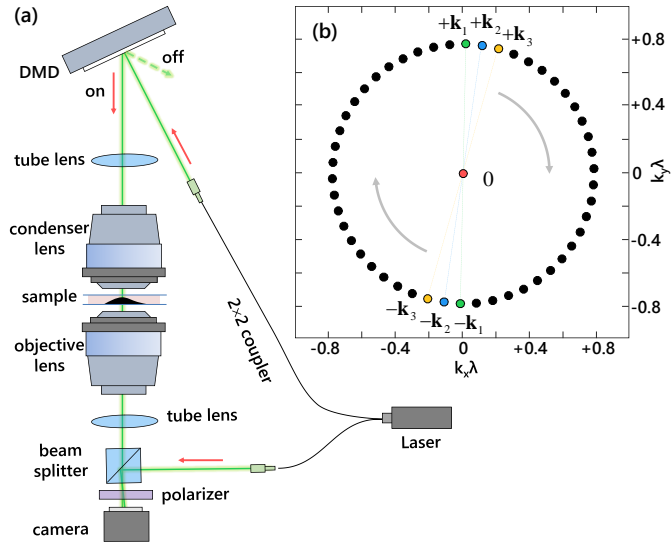

**Fig. 2.** Experimental setup. (a) Optical setup based on Mach-Zehnder interferometry. The structured illumination displayed onto a DMD is projected onto a sample plane. (b) The schematic of 51 spatial frequency components addressed by using the patterns in Eq. (3). $k_x$ and $k_y$ are the $x$- and $y$-directional components of the wavevectors, respectively. The gray arrow indicates the direction of circular scanning.

The sample beam is reflected by a DMD (DLP LightCrafter 3000, Texas Instruments Inc., TX, USA) in order to control the wavefront of the incident beam. The DMD displays 8-bit sinusoidal patterns as in Eq. (1). Then, the patterned beam was projected onto a sample plane, after being demagnified by a tube lens ($f$ = 200 mm) and a high numerical aperture (NA) condenser lens (NA = 1.2, water immersion, UPLSAPO 60XW, 60×, Olympus Inc., Japan). It is noteworthy that the present technique does not require the use of any additional iris or aperture for filtering unwanted diffraction patterns.

Then the beam diffracted from a sample was collected by a high NA objective lens (NA = 1.2, water immersion, UPLSAPO 60XW, 60×, Olympus Inc., Japan) and projected onto an image plane via a tube lens ($f$ = 180 mm). At the image plane, the diffracted and the reference beam interfered, and generated a spatially modulated hologram. The hologram was recorded by a CCD camera (FL3-U3-13Y3M-C, FLIR Systems, Inc., OR, USA). The complex optical fields of the sample $X_\mathbf{k}^\varphi$ which was illuminated with a specific sinusoidal pattern, were retrieved from the measured holograms via a field retrieval algorithm [33, 34].

In order to retrieve individual scattered fields $U_\mathbf{k}$, we performed the following decomposition process. Because sinusoidal intensity patterns share the unmodulated (normal) illumination or DC term [Eq. (2)], we first measured $U_0$ separately. Then, for each wavevector $\mathbf{k}$, two additional measurements with $\varphi = 0, \pi/2$ were performed to extract $U_{+\mathbf{k}}$ and $U_{-\mathbf{k}}$. Therefore, the present method requires $2N+1$ fields measurements for $N$ sinusoidal patterns, which can be described as,

$$\begin{aligned}
X_0^0 &= 2U_0, \\
X_{\mathbf{k}_1}^0 &= U_0 + \tfrac{1}{2}U_{+\mathbf{k}_1} + \tfrac{1}{2}U_{-\mathbf{k}_1}, \\
X_{\mathbf{k}_1}^{\pi/2} &= U_0 + \tfrac{i}{2}U_{+\mathbf{k}_1} - \tfrac{i}{2}U_{-\mathbf{k}_1}, \\
&\vdots \\
X_{\mathbf{k}_N}^0 &= U_0 + \tfrac{1}{2}U_{+\mathbf{k}_N} + \tfrac{1}{2}U_{-\mathbf{k}_N}, \\
X_{\mathbf{k}_N}^{\pi/2} &= U_0 + \tfrac{i}{2}U_{+\mathbf{k}_N} - \tfrac{i}{2}U_{-\mathbf{k}_N}.
\end{aligned} \quad (3)$$

By solving Eq. (3), a total of $2N+1$ scattered fields components ($U_0$, $U_{+k_1}$, $U_{-k_1}$, ..., $U_{+k_N}$, $U_{-k_N}$) were retrieved. In the following experiments, we performed circular scanning composed of 51 plane wave components [Fig. 2(b)].

In order to validate the proposed method, we experimentally performed the retrieval of scattered field components (Fig. 3). From the experimentally measured hologram obtained with the time-multiplexed structured illumination [Fig. 3(a)], the scattered fields are retrieved after decomposition [Fig. 3(b)]. Please note that in order to decompose these three scattered fields, we measured three holograms: $X_0^0, X_\mathbf{k}^0, X_\mathbf{k}^{\pi/2}$.

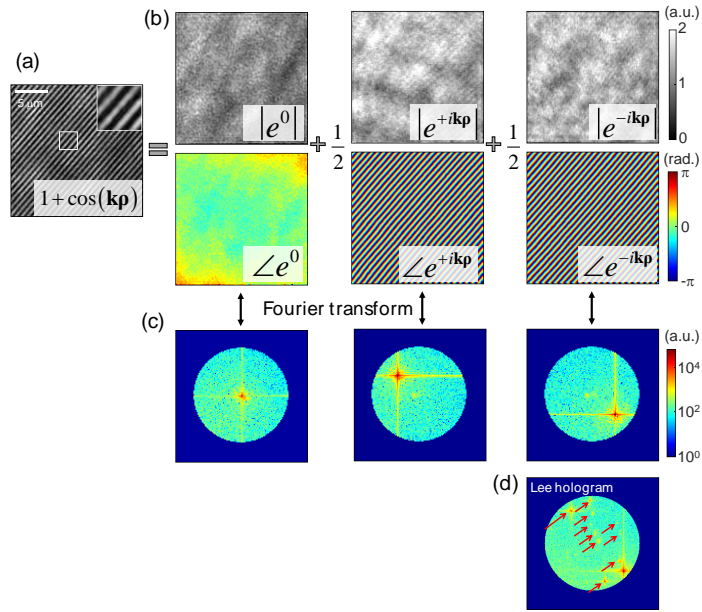

**Fig. 3**. Experimental measurements of the time-multiplexed structured illumination. (a) Hologram measured with the time-multiplexed structured illumination. (b) Retrieved amplitude and phase maps after decomposition. (c) Corresponding Fourier spectra maps, which clearly show each spatial frequency component. (d) Fourier spectra map obtained with the method using Lee holograms as in Ref. 27. There are several unwanted diffraction patterns, as indicated by the red arrows.

The corresponding Fourier spectra maps show individual spatial frequency components without diffraction noise. In contrast, the Fourier spectra of the optical field obtained with the previous method using Lee holograms [Fig. 3(d)] show several unwanted diffraction noise patterns (indicated with the red arrows), which significantly deteriorate the image quality.

From the retrieved scattered field components, the 3-D RI distribution of the sample was reconstructed by using an ODT algorithm [1, 5]. Each 2-D Fourier spectrum of a complex optical field was mapped onto the corresponding Ewald surface in the 3-D Fourier space according to the Fourier diffraction theorem [1, 35]. Then, the 3-D RI distribution of the sample was reconstructed by applying the 3-D inverse Fourier transformation to the 3-D Fourier space. Detailed procedures and a MatLab code can be found elsewhere [5].

In order to validate the feasibility of the present method, we measured the 3-D RI distributions of various samples including colloidal and biological samples. We first measured a silica bead with a diameter of 5 μm (44054, $n$ = 1.4607 at $\lambda$ = 532 nm, Sigma-Aldrich Inc., MO, USA) immersed in 55% (w/w) sucrose solution ($n$ = 1.43) (Fig. 4).

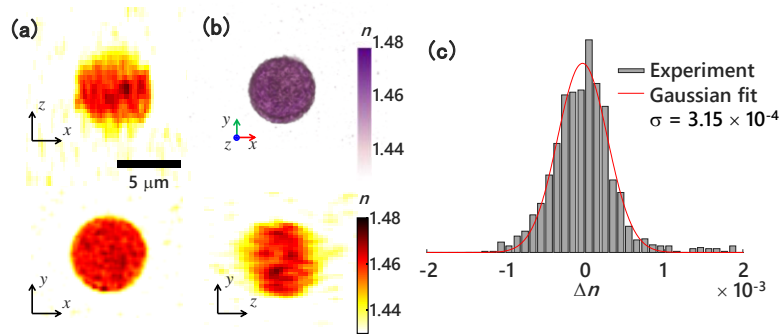

**Fig. 4.** (a) Cross-sectional slices and (b) 3-D rendered images of the reconstructed 3-D RI distribution of a silica bead with a diameter of 5 μm. (c) The distribution of RI contrast values in the background region of the tomogram in (a). The solid red line indicates a Gaussian fit with the standard deviation of $3.15 \times 10^{-4}$.

As shown in Figs. 4(a)–(b), the shape and RI value of the reconstructed tomogram of the silica bead are in good agreement with the manufacturer's specification. In addition, the reconstructed RI tomogram exhibits high RI sensitivity. As shown in Fig. 4(c), the standard deviation of the RI contrast values in the background region was $\sigma_{\Delta n} = 3.15 \times 10^{-4}$. This high RI sensitivity is attributed to the elimination of unwanted diffraction noise as a result of using the time-multiplexed structured illumination.

To further demonstrate the applicability of the method, we measured the 3-D RI distributions of live biological cells (Fig. 5). The reconstructed 3-D RI distribution of a red blood cell (RBC) is presented in Fig. 5(a). Blood from a healthy donor was diluted in a phosphate buffered saline solution ($n$ = 1.337 at $\lambda$ = 532 nm, Welgene Inc., Korea), and individual RBCs in the diluted blood were measured using the present system. The measured 3-D RI map of the RBC clear shows the characteristic discoid shape, as in the cross-sectional slices [the left panel of Fig. 5(a)] and a 3-D rendered image [the right panel of Fig. 5(a)]. From the measurements of 10 RBCs, we calculated quantitative information including the biochemical and morphological parameters of individual RBCs. Cellular volume, hemoglobin (Hb) concentration, and Hb contents were calculated to be $89.47 \pm 7.88$ fL, $33.67 \pm 0.69$ g/dL, and $30.10 \pm 2.40$ pg, respectively, which in the normal physiological condition [6]. Furthermore, the 3-D RI distribution of a HeLa cell was also measured [Fig. 5(b)], which shows more complicated subcellular structures.

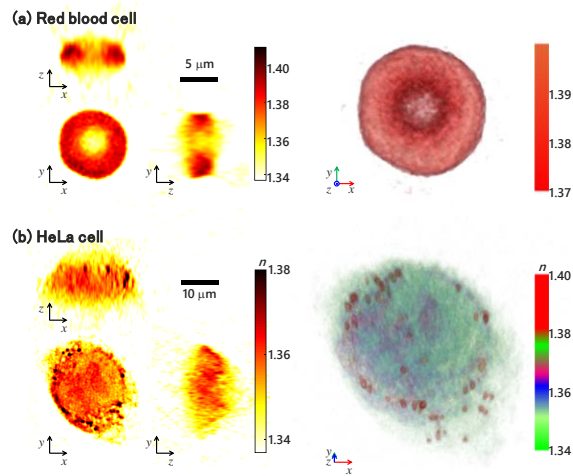

**Fig. 5.** Cross-sectional slices (left panel) and 3-D rendered images of the reconstructed 3-D RI distribution of (a) a red blood cell, and (b) a HeLa cell.

In conclusion, we proposed and experimentally demonstrated a novel illumination control technique for 3-D RI tomograms using a DMD and time-multiplexed structured illumination. The present illumination method effectively controls the wavefronts so that scattered fields from a sample with various spatial frequencies were precisely retrieved. The time-multiplexing projection of sinusoidal intensity patterns does not produce unwanted diffraction noise, compared to the previous method using binary Lee holograms. We demonstrated the performance of the present method by measuring various types of microscopic samples including a silica bead, a human RBC, and a HeLa cell. The 3-D RI tomograms of the samples were precisely measured with minimal diffraction noise; the reconstructed tomograms showed high RI sensitivity, of $\sigma_{\Delta n} = 3.15 \times 10^{-4}$.

We envision that the present technique can be applied to various imaging modalities that require precise illumination control with the use of a DMD, such as synthetic aperture holographic microscopy [36, 37] and structured illumination fluorescent microscopy [38]. Since a DMD provides ultra-fast control of an illumination beam with a repetition rate of up to tens of kHz, we anticipate that the use of the present technique with various imaging modalities will open a new avenue for the investigation of various imaging sciences.

## Acknowledgement

This work was supported by Tomocube, and the National Research Foundation of Korea (2015R1A3A2066550, 2014K1A3A1A09063027, 2014M3C1A3052567) and Innopolis foundation (A2015DD126). Prof. Park has financial interests in Tomocube Inc., a company that commercializes ODT.

## References

1. E. Wolf, "Three-dimensional structure determination of semi-transparent objects from holographic data," Optics Communications **1**, 153-156 (1969).
2. O. Haeberle, K. Belkebir, H. Giovaninni, and A. Sentenac, "Tomographic diffractive microscopy: basics, techniques and perspectives," Journal of Modern Optics **57**, 686-699 (2010).
3. K. Lee, K. Kim, J. Jung, J. H. Heo, S. Cho, S. Lee, G. Chang, Y. J. Jo, H. Park, and Y. K. Park, "Quantitative phase imaging techniques for the study of cell pathophysiology: from principles to applications," Sensors **13**, 4170-4191 (2013).
4. K. Kim, J. Yoon, S. Shin, S. Lee, S.-A. Yang, and Y. Park, "Optical diffraction tomography techniques for the study of cell pathophysiology," Journal of Biomedical Photonics & Engineering **2**, 020201 (2016).
5. K. Kim, H.-O. Yoon, M. Diez-Silva, M. Dao, R. Dasari, and Y.-K. Park, "High-resolution three-dimensional imaging of red blood cells parasitized by *Plasmodium falciparum* and *in situ* hemozoin crystals using optical diffraction tomography," Journal of Biomedical Optics **19**, 011005-011012 (2014).
6. Y. Kim, H. Shim, K. Kim, H. Park, J. H. Heo, J. Yoon, C. Choi, S. Jang, and Y. Park, "Common-path diffraction optical tomography for investigation of three-dimensional structures and dynamics of biological cells," Optics Express **22**, 10398-10407 (2014).
7. T. Kim, R. Zhou, M. Mir, S. D. Babacan, P. S. Carney, L. L. Goddard, and G. Popescu, "White-light diffraction tomography of unlabelled live cells," Nature Photonics **8**, 256-263 (2014).
8. H. Park, T. Ahn, K. Kim, S. Lee, S. Y. Kook, D. Lee, I. B. Suh, S. Na, and Y. Park, "Three-dimensional refractive index tomograms and deformability of individual human red blood cells from cord blood of newborn infants and maternal blood," Journal of Biomedical Optics **20**, 111208 (2015).
9. S. Y. Lee, H. J. Park, C. Best-Popescu, S. Jang, and Y. K. Park, "The Effects of Ethanol on the Morphological and Biochemical Properties of Individual Human Red Blood Cells," PLoS One **10**, e0145327 (2015).
10. K. Kim, K. Choe, I. Park, P. Kim, and Y. Park, "Holographic intravital microscopy for 2-D and 3-D imaging intact circulating blood cells in microcapillaries of live mice," Scientific Reports **6**, 33084 (2016).
11. J. Yoon, K. Kim, H. Park, C. Choi, S. Jang, and Y. Park, "Label-free characterization of white blood cells by measuring 3D refractive index maps," Biomedical Optics Express **6**, 3865-3875 (2015).
12. K. Kim, J. Yoon, and Y. Park, "Simultaneous 3D visualization and position tracking of optically trapped particles using optical diffraction tomography," Optica **2**, 343-346 (2015).
13. Y. Cotte, F. Toy, P. Jourdain, N. Pavillon, D. Boss, P. Magistretti, P. Marquet, and C. Depeursinge, "Marker-free phase nanoscopy," Nature Photonics **7**, 113-117 (2013).
14. S.-A. Yang, J. Yoon, K. Kim, and Y. Park, "Measurements of morphological and biochemical alterations in individual neuron cells associated with early neurotoxic effects in Parkinson's disease," bioRxiv preprint, 080937 (2016).
15. R. Chandramohanadas, Y. Park, L. Lui, A. Li, D. Quinn, K. Liew, M. Diez-Silva, Y. Sung, M. Dao, C. T. Lim, P. R. Preiser, and S. Suresh, "Biophysics of malarial parasite exit from infected erythrocytes," PLoS One **6**, e20869 (2011).
16. H. Park, S. H. Hong, K. Kim, S. H. Cho, W. J. Lee, Y. Kim, S. E. Lee, and Y. Park, "Characterizations of individual mouse red blood cells parasitized by Babesia microti using 3-D holographic microscopy," Scientific Reports **5**, 10827 (2015).
17. W.-C. Hsu, J.-W. Su, C.-C. Chang, and K.-B. Sung, "Investigating the backscattering characteristics of individual normal and cancerous cells based on experimentally determined three-dimensional refractive index distributions," Proc. of SPIE **8553**, 85531O (2012).
18. K. Kim, Z. Yaqoob, K. Lee, J. W. Kang, Y. Choi, P. Hosseini, P. T. C. So, and Y. Park, "Diffraction optical tomography using a quantitative phase imaging unit," Optics Letters **39**, 6935-6938 (2014).
19. K. Kim, J. Yoon, and Y. Park, "Large-scale optical diffraction tomography for inspection of optical plastic lenses," Optics Letters **41**, 934-937 (2016).
20. V. Lauer, "New approach to optical diffraction tomography yielding a vector equation of diffraction tomography and a novel tomographic microscope," Journal of Microscopy **205**, 165-176 (2002).
21. Y. Sung, W. Choi, C. Fang-Yen, K. Badizadegan, R. R. Dasari, and M. S. Feld, "Optical diffraction tomography for high resolution live cell imaging," Optics Express **17**, 266-277 (2009).
22. K. Kim, K. S. Kim, H. Park, J. C. Ye, and Y. Park, "Real-time visualization of 3-D dynamic microscopic objects using optical diffraction tomography," Optics Express **21**, 32269-32278 (2013).
23. A. Kuś, W. Krauze, and M. Kujawińska, "Active limited-angle tomographic phase microscope," Journal of Biomedical Optics **20**, 111216 (2015).
24. F. Charriere, A. Marian, F. Montfort, J. Kuehn, T. Colomb, E. Cuche, P. Marquet, and C. Depeursinge, "Cell refractive index tomography by digital holographic microscopy," Optics Letters **31**, 178-180 (2006).

25. A. Kuś, M. Dudek, B. Kemper, M. Kujawińska, and A. Vollmer, "Tomographic phase microscopy of living three-dimensional cell cultures," Journal of Biomedical Optics **19**, 046009 (2014).
26. P. Müller, M. Schürmann, C. J. Chan, and J. Guck, "Single-cell diffraction tomography with optofluidic rotation about a tilted axis," Proc. of SPIE **9548**, 95480U (2015).
27. S. Shin, K. Kim, J. Yoon, and Y. Park, "Active illumination using a digital micromirror device for quantitative phase imaging," Optics Letters **40**, 5407-5410 (2015).
28. S. Shin, K. Kim, T. Kim, J. Yoon, K. Hong, J. Park, and Y. Park, "Optical diffraction tomography using a digital micromirror device for stable measurements of 4-D refractive index tomography of cells," Proc. of SPIE **9718**, 971814 (2016).
29. W.-H. Lee, "Binary computer-generated holograms," Applied Optics **18**, 3661-3669 (1979).
30. D. B. Conkey, A. M. Caravaca-Aguirre, and R. Piestun, "High-speed scattering medium characterization with application to focusing light through turbid media," Optics Express **20**, 1733-1740 (2012).
31. M. G. L. Gustafsson, "Surpassing the lateral resolution limit by a factor of two using structured illumination microscopy," Journal of Microscopy **198**, 82-87 (2000).
32. I. Yamaguchi and T. Zhang, "Phase-shifting digital holography," Optics Letters **22**, 1268-1270 (1997).
33. M. Takeda, H. Ina, and S. Kobayashi, "Fourier-transform method of fringe-pattern analysis for computer-based topography and interferometry," Journal of the Optical Society of America **72**, 156-160 (1982).
34. S. K. Debnath and Y. Park, "Real-time quantitative phase imaging with a spatial phase-shifting algorithm," Optics Letters **36**, 4677-4679 (2011).
35. A. C. Kak and M. Slaney, *Principles of computerized tomographic imaging*, Classics in applied mathematics (Society for Industrial and Applied Mathematics, Philadelphia, 2001).
36. S. Alexandrov, T. Hillman, T. Gutzler, and D. Sampson, "Synthetic aperture Fourier holographic optical microscopy," Physical Review Letters **97**, 168102 (2006).
37. K. Lee, H. D. Kim, K. Kim, Y. Kim, T. R. Hillman, B. Min, and Y. Park, "Synthetic Fourier transform light scattering," Optics Express **21**, 22453-22463 (2013).
38. L. Shao, P. Kner, E. H. Rego, and M. G. L. Gustafsson, "Super-resolution 3D microscopy of live whole cells using structured illumination," Nature Methods **8**, 1044-1046 (2011).